# Carbon-atom wires produced by nanosecond pulsed laser deposition in a background gas

C.S. Casari*, C.S. Giannuzzi, V. Russo

Department of Energy, Politecnico di Milano

Via Ponzio 34/3 20133 Milan, Italy

Abstract

Wires of *sp*-hybridized carbon atoms are attracting interest for both fundamental aspects of carbon science and for their appealing functional properties. The synthesis by physical vapor deposition has been reported to provide sp-rich carbon films but still needs to be further developed and understood in detail. Here the synthesis of carbon-atom wires (CAWs) has been achieved by nanosecond pulsed laser deposition (PLD) expoliting the strong out-of-equilibrium conditions occurring when the ablation plasma is confined in a background gas. Surface Enhnaced Raman scattering (SERS) spectra of deposited films indicates that CAWs are mixed with a mainly $sp^2$ amorphous carbon in a $sp$-$sp^2$ hybrid material. Optimal conditions for the deposition of *sp*-carbon phase have been investigated by changing deposition parameters thus suggesting basic mechanisms of carbon wires formation. Our proof-of-concept may open new perspectives for the targeted fabrication of CAWs and $sp$-$sp^2$ structures.

*Corresponding author: Tel. +390223996331 E-mail: carlo.casari@polimi.it  (Carlo S. Casari)

# 1. Introduction

In the family of carbon nanostructures comprising fullerenes as a 0D system, nanotubes as quasi-1D and graphene as the parent 2D material, carbon-atom wires (CAWs) represent the ultimate 1D systems [1,2]. CAWs, formed by *sp*-hybridized carbon atoms, are revealing a great potential in terms of electronic, optical and mechanical properties, as recently shown by experimental observations and theoretical predictions [2-7]. In the model CAW, a strong structure-property relationship, originating from electron conjugation effects, leads to metallic character of cumulene (i.e. wire with all double bonds) and semiconducting one of polyyne (i.e. alternate single-triple bond structure). In real systems, a wide tunability of electronic and optical properties is thus accessible by playing with finite-length and termination-induced effects [2]. The formation of *sp*-carbon linear structures is a relevant step for understanding fundamental mechanisms of carbon atoms coagulation, at the base of the growth of larger carbon nanostructures including fullerenes, novel allotropes and *sp*-$sp^2$ hybrid structures (e.g. graphyne). In astrophysics and astrochemistry a fundamental question regards the nature of the carriers of diffuse interstellar bands observed in astronomical spectroscopy. Small carbon clusters such as linear chains or rings of *sp*-hybridized carbon and up to fullerenes are considered as candidates and laboratory investigations are widely used to study molecular and carbon aggregates in the gas phase with the aim of assigning observed bands to specific species, as recently shown by P. Maier and co-workers [8,9]. Despite the interest in *sp*-carbon, many aspects of the formation mechanisms are still open and much experimental work is needed to unveil optimal conditions for *sp*-carbon growth and to develop effective production techniques.

A number of techniques have already shown the capability to produce CAWs in different forms (isolated structures, in liquids, in solid matrices and in thin films) [10-17]. Physical methods are mainly based on a bottom-up approach consisting in the production of a carbon vapour followed by a rapid quenching (in gas or liquid phase) to obtain structures formed in strongly out-of-equilibrium conditions[16-18]. Liquid-based methods such as the submerged arc discharge or the pulsed laser ablation in liquids allow to produce isolated CAWs (mainly polyynes) in solution [15, 19,20]. In the gas phase, low energy cluster beam deposition has demonstrated the capability to produce a sp-rich pure carbon material in which wires and amorphous $sp^2$ carbon phase coexist [17, 21,22]. For this technique, dedicated experiments have reveled that *sp*-carbon is already present in carbon clusters in the gas phase [23]. Recently Taguchi et al. have investigated the formation of Hydrogen-terminated polyynes by laser irradiating a graphite target in the presence of propane and collecting the species in a liquid solvent [24].

Nanosecond pulsed laser deposition (PLD) is a versatile technique capable to produce carbon films with a control of the $sp^2/sp^3$ content [25]. The possibility to ablate virtually any type of target material in different conditions (e.g. vacuum or background atmosphere) gives access to a fine control of the deposited material (e.g. morphology, stoichiometry, doping) [26]. Despite the wide use of ns PLD for the production of carbon films, the presence of carbon-atom wires has not been reported yet. In this framework some papers have claimed that the formation of $sp$-carbon is limited only to the use of fs laser pulses in vacuum [27].

One of the most assessed techniques for detecting $sp$-carbon is Raman spectroscopy [3]. The high sensitivity to the local bond organization and the peculiar Raman features well separated from those of other forms of carbon makes Raman spectroscopy particularly suitable for investigating carbon-atom wires [3, 17, 28-30]. In particular Surface Enhanced Raman Spectroscopy (SERS), exploiting enhancement effects of surface plasmons of metallic nanoparticles, is able to provide unambiguous detection both in liquids and in the solid state even at low concentration of sample material [3, 31-34].

In this work we show that sp carbon-atom wires embeeded in a mainly $sp^2$ amorphous network can be produced by conventional ns pulsed laser deposition in the presence of Ar as a background gas. Carbon films have been deposited on SERS-active substrates at room temperature and by means of SERS we were able to investigate in detail the ablation conditions that maximize the presence of $sp$-carbon. Our presented approach opens the possibility to unveil the formation mechanisms of $sp$-carbon, that is a crucial step towards the understanding of the growth of carbon nanostructures from small clusters to fullerenes and nanotubes. The synthesis of CAWs embedded in an amorphous carbon is attractive for the development of novel $sp$-$sp^2$ carbon hybrids consisting in $sp^2$ domains connected by linear wires.

**2. Experimental details**

PLD of carbon films was performed at room temperature by focusing a frequency doubled lamp-pumped Q-switched Nd:YAG pulsed laser (λ = 532 nm, pulse duration 7-9 ns, 10 Hz repetition rate) on a graphite target (purity 99.999%) in a vacuum chamber. The laser energy (250 mJ) and the spot size on the target were adjusted to have an energy density on the target (i.e. fluence) of about 2.7 J/cm$^2$. Ablation was performed in the presence of Argon with a background pressure ($P_{Ar}$) varying in the 10-500 Pa range and a target-to-substrate distance ($d_{ts}$) in the 25-110 mm range. Deposition time was adjusted according to the pressure and $d_{ts}$, which strongly affect the deposition rate (typical values are between 1 and 30 minutes). The vacuum chamber is equipped with a scroll pump and a turbomolecular pump with a base pressure of about $10^{-3}$ Pa. Backgorund gas pressure is tuned by means of flow mass controllers and capacitance pressure gauges. Time-integrated pictures of the plume shape were taken by a camera through a viewport. Surface Enhance Raman Scattering SERS spectra were acquired with a Renishaw InVia micro-Raman spectrometer using the 514.5 nm wavelength of an $Ar^+$ laser and a Peltier-cooled CCD camera, with a spectral resolution of about 3 $cm^{-1}$. SERS was accomplished by directly depositing carbon on SERS-active substrates consisting of Ag nanoparticles grown on Si substrates by thermal evaporation. The equivalent Ag thickness, as monitored by a quartz microbalance, was adjusted at about 3 nm in order to have a good matching between the resonance plasmon peak and the excitation laser line used for Raman (i.e. 514.5 nm), as measured by UV-vis absorption spectroscopy (not shown). As prepared SERS-active substrates were accurately cleaned to avoid SERS signals coming from contaminations. In particular a cleaning procedure has been adopted, just before PLD depositon, consisting in cleaning in isopropanol and ultrasonic bath for a few minutes followed by drying under gentle flux of pure nitrogen. Such procedure ensures that no features in the 1800-2200 $cm^{-1}$ spectral range (i.e. region of *sp*-carbon) are present in the spectra of pristine SERS-active substrates due to ambient contaminations. A weak contribution, consisitng of sharp peaks in the region of amorphous $sp^2$ carbon (1200-1600 $cm^{-1}$), is always present possibly due to carbon-based molecular species present in ambient air [35]. The cleaning procedure do not remove or modify Ag nanoparticles deposited on the substrate, as checked by SEM and SERS-activity.

Laser power on the PLD samples was carefully selected (1mW) to prevent laser-induced damage and graphitization. Scanning electron microscopy (SEM) images were acquired with a Zeiss Supra 40 field emission SEM.

## 3. Results and discussion

SERS spectra of carbon films deposited by PLD at different Ar background gas pressure are shown in figure 1. Two main features characterize the spectra: one in the 1200-1600 $cm^{-1}$ range and the other in the 1800-2200 $cm^{-1}$. The first is a band comprising G and D modes characteristic of $sp^2$-hybridized carbon, the latter indicates the presence of *sp*-carbon wires. The G mode at 1580 $cm^{-1}$ is characteristic of CC streching mode in graphite while the D peak appearing at about 1350 $cm^{-1}$ is related to breathing mode of hexagonal rings and is activated by structural disorder [36]. The *sp*-carbon features are due to stretching modes of the chain. The specific frequency is strongly related to the length and structure of the wire [3].

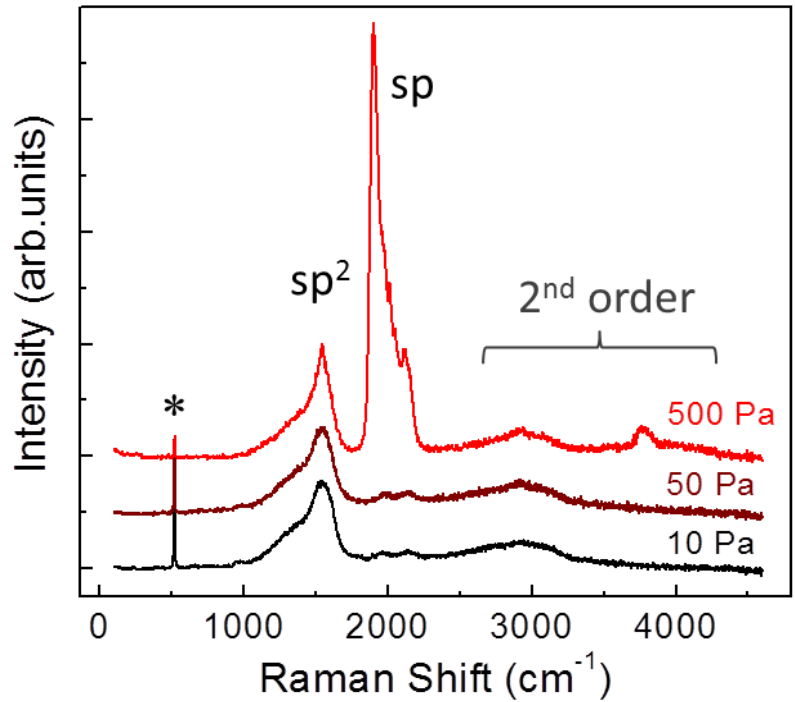

*Figure 1. Extended SERS spectra of carbon films deposited by PLD at different Ar gas pressure (10, 50 and 500 Pa). The target-to-substrate distance is kept constant ($d_{ts}$ =50 mm). Spectra are normalized with respect to the G band intensity for the sake of comparison. The signal from silicon coming from the SERS-active substrate is labeled with a star.*

The large width of G and D peaks and their superposition in a broad band indicates amorphous $sp^2$ carbon phase. Remarkably, the signal of *sp*-carbon is very weak at low Ar pressure ($P_{Ar}$ =10 and 50 Pa) and increases consistently for $P_{Ar}$ =500 Pa finally showing intensity more than 2.5 times higher than that of the G peak. In addition, the *sp*-carbon band is structured in different contributions, the most intense ones being at about 1900 cm$^{-1}$ and 2130 cm$^{-1}$. All the features above 2500 cm$^{-1}$ frequency are due to second order Raman scattering. The large band in the 2500-3500 cm$^{-1}$ region is the second order of $sp^2$ carbon phase which is typically broad and unresolved in the case of amorphous carbon. The band appearing at 3770 cm$^{-1}$ only in the spectrum of sample deposited at 500 Pa can be associated to second order Raman band of *sp*-carbon, as the overtone of the peak at 1900 cm$^{-1}$. Second order Raman contribution is usually less intense than the correspondent first order and it appears only when the first order is very intense, as in the case of 500 Pa Ar.

As a general trend we observe that a high pressure is required to maximize the intensity of the *sp*-carbon peak in the SERS spectra. However, the fraction of sp carbon phase in the whole sample is difficult to estimate. First, we underline that in our solid-SERS experiments, the signal is mainly coming from wires that are close or linked to metal nanoparticles covering the substrate surface (i.e. chemical SERS effect). Second, even assuming that the presence of *sp*-carbon is uniform in the sample volume, the SERS enhancement effect of sp and *sp*$^2$ carbon phases can be different as well as the Raman scattering cross sections. To this respect the Raman cross section of sp carbon can be between 1.2 and 3 times larger than the cross section of *sp*$^2$ carbon, as evaluated from x-ray absorption spectra [37].

With the aim of understanding the optimal conditions for the formation of *sp*-carbon wires we investigated also the effect of changing the target-to-substrate distance ($d_{ts}$). By fixing the Ar pressure we have deposited at different $d_{ts}$, the spectra are shown in figure 2 for 50 and 500 Pa Ar pressure. As a general trend we observe that the *sp*-carbon signal increases when increasing $d_{ts}$. In the case of 50 Pa Ar, *sp*-carbon band is negligible for $d_{ts}$ = 50 mm and increases consistently with $d_{ts}$ finally overcoming the intensity of the G peak for $d_{ts}$ = 110 mm. At 500 Pa Ar the effect is similar to the one observed at 50 Pa, *sp*-carbon signal increases with $d_{ts}$, though for different values of $d_{ts}$ and with an overall higher intensity of *sp*-carbon.

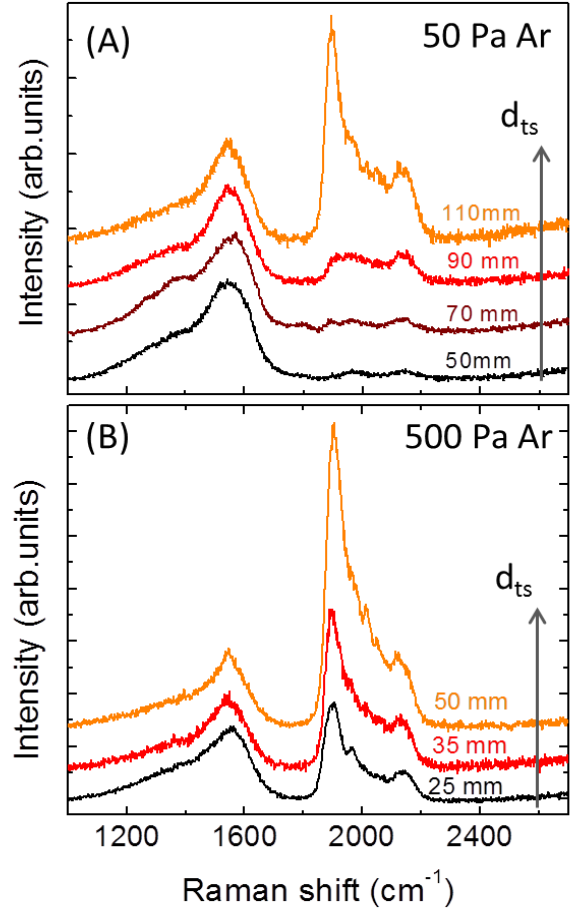

*Figure 2. SERS spectra as a function of the target-to-substrate distance $d_{ts}$ for a fixed Ar gas pressure of (A): 50 Pa and (B): 500 Pa. The spectra are normalized with respect to the G band intensity.*

The strong dependence of the *sp*-carbon peak on $d_{ts}$ that is emerging from our results is also reflected in a different film morphology, as observed by SEM images shown in figure 3. At a fixed pressure of 50 Pa Ar the film morphology evolves from a more compact columnar-like to open-porous and foam-like when increasing $d_{ts}$ from 50 mm to 110 mm. When fixing the pressure at 500 Pa this general behavior is mantained even if for different values of $d_{ts}$. As what happens when increasing $d_{ts}$ at constant pressure, we observe morphology change from compact to porous also when pressure is increased at a fixed $d_{ts}$, as evident form the comparison of figure 3-a and 3-d. The background gas pressure act as a confining agent on the plasma. Confinement leads to an increase in visible emission by the plasma which reaches, while expanding, a well defined plume shape with visible boundaries. The time integrated plume shape is characterized by a plume length ($l_p$) that is the maximum distance from the target reached by the plasma boundary (i.e. shock front). Pressure and distance $d_{ts}$ have been already demonstrated to play a lead role in the PLD process [38]. The non trivial interplay can be roughly condensed in one simple experimental parameter L, defined as the ratio between the target-to-substrate distance $d_{ts}$ and the visible plume length $l_p$ (L= $d_{ts}/ l_p$). Since $l_p$ is the time integrated plume length, it defines the maximum position reached by the shock front during plasma plume expansion. Hence the L parameter indicates the relative position of the shock front with respect to the substrate (see insets of figure 3). For a number of materials including metals and oxides we have already demonstrated that L can be used as a guide for selecting the growth mode of deposited films: L<1 leads to compact films mainly grown atom-by-atom, while when L>1 clustering phenomena and diffusion of ablated species in the surrounding gas lead to low energy cluster deposition and growth of cluster-assembled nanoporous materials (including multiscale hierarchically structured films) [26].

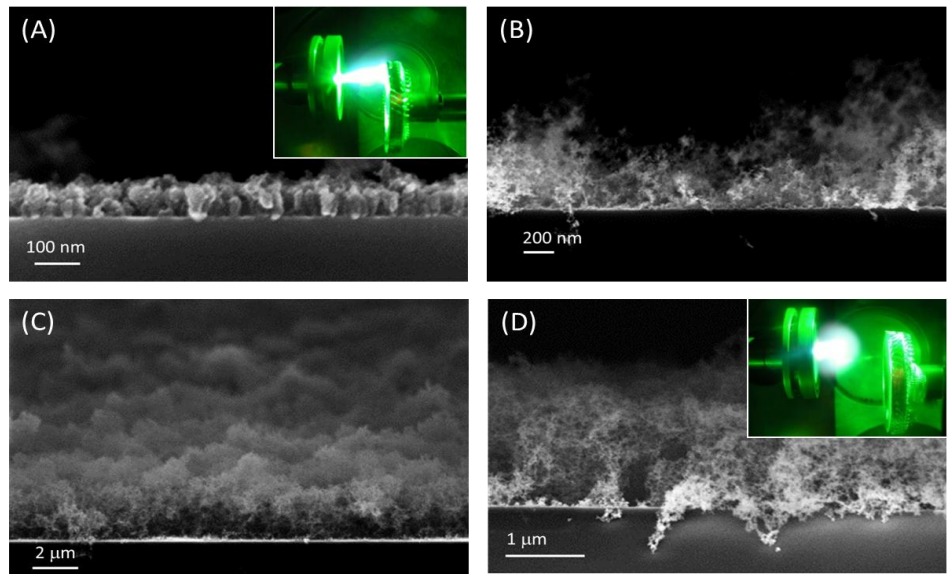

*Figure 3. Cross sectional SEM images of carbon PLD films on SERS-active substrates. (Top line): Ar gas pressure of 50Pa and target-to-substrate distance $d_{ts}$ = 50 mm (A) and $d_{ts}$ = 110 mm (B). (Bottom line): Ar gas pressure of 500Pa and target-to-substrate distance $d_{ts}$ = 35 mm (C) and $d_{ts}$ = 50 mm (D). Pictures of the plasma plume are reported in the inset.*

This is of great relevance for the deposition process because the plume shock front represents the boundary between the inner part of the plume and the outer region occupied by the surrounding gas. The shock front is a region characterized by strong gradients of temperature, velocity and pressure, where clustering of the ablated species can occur in out-of-equilibrium conditions [39-42].

Moreover when ablated species can diffuse in the background gas before being deposited, their kinetic energy is strongly reduced down to less than 1eV/atom which corresponds to a low energy cluster deposition regime. In this case clusters experience a soft landing on the substrate with negligible fragmentation thus preserving the original structure formed during the plume expansion.[43]

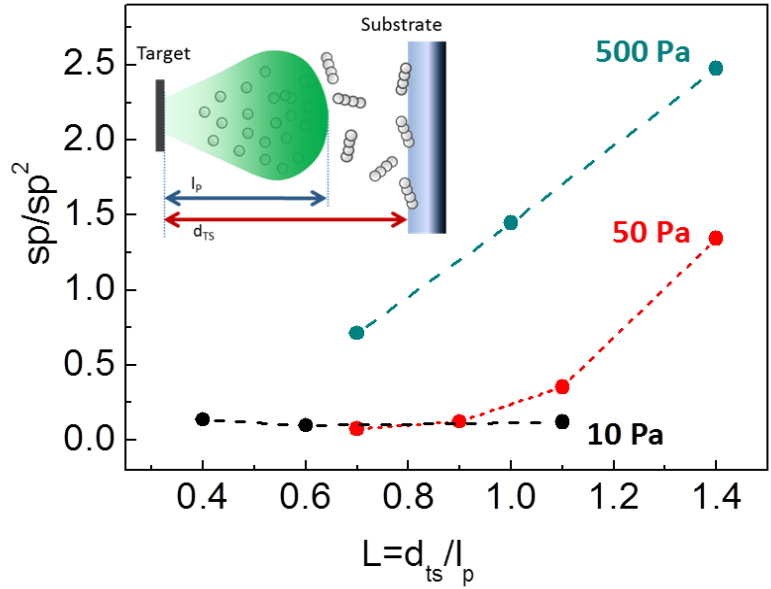

*Figure 4. Plot of the* sp *to sp$^2$ intensity ratio (as measured from spectra of figure 2) as a function of the non-dimensional distance L for 10, 50 and 500 Pa Ar. The optimum condition for sp-carbon wire formation (L>1) is shown in the inset.*

There is a general agreement about the idea that formation of *sp*-carbon wires occurs in stronlgy out-of-equilibrium conditions. Thus the L parameter is likely to play a key role in selecting the suitable conditions for wire formation in the plasma plume. With reference to the PLD conditions investigated in this work, at 50 Pa Ar the plume length is about 50 mm while it reduces to about 35 mm at 500 Pa. In these plume conditions $d_{ts}$ was selected to have the substrate crossing (i.e. L<1) or not crossing (i.e. L>1) the plasma plume. To better clarify the role of L in the formation process we plotted the *sp*/ *sp$^2$* intensity ratio (intensity calculated by the total area under the corresponding band) as a function of L for different Ar pressures (see figure 4). At 10 Pa Ar (spectra not shown) no relevant signal of sp carbon wires is observed in the investigated range (0.4<L<1.1). At 50 Pa a high *sp*-carbon signal is detected when crossing the condition L=1, which corresponds to setting the visible plume tangent to the substrate. At 500 Pa the trend appears to be shifted at lower L values and the same condition found at 50 Pa for L=1.4 now occurs for L=1. The maximum sp signal is observed at 500 Pa Ar for L=1.4 that is more than 2.5 times the intensity of the G peak. It is clear that high pressure is able to provide suitable conditions for CAWs formation. Then, by positioning the substrate far from the shock front (i.e. high L values) a low energy deposition regime allows for a soft landing on the substrate thus preventing reorganization in more stable phase. The slowdown effect of ablated species has a direct consequence in the deposition rate which drops down dramatically at large values of L parameter.

Our results indicate that CAWs form in the gas phase where high pressure and temperature gradients produce strong out-of-equilibrium conditions (i.e. at the plume shock front). Some works have modeled shock waves propagation in 500 Pa of ambient gas after laser ablation of graphite showing pressure, temparture and velocity values up to 1 MPa, 8000 m/s and $10^5$ K, respectively concentrated in a region of about 0.1 mm from the target [44]. Furthermore, experimental works focusing on the mechanisms of ablation of graphite in the presence of a backgorund gas have outlined the formation of C2 dimers with kinetic energy of several eV. For instance, the collisional heating is increased when increasing the backgound gas pressure reaching 12000 K at a distance of 8 mm from the target whe using about 70 Pa N2 [45]. These conditions seem to meet the requirements for the formation of *sp*-carbon according to carbon phase diagrams proposed by diferent authors. Recently, the conditions for *sp*-carbon formation by thermal decomposition of $sp^2$ carbon in a discharge plasma have been simulated by molecular dynamics [46]. Even though the processes in laser ablated plasma are different from those simulated, some general indications can be extracted. First, the size distribution in the hot region is constituted by atoms, dimers and wires up to about 30 carbon atoms. Second, the main mechanism for optimal production seems to be related to a proper balance between a sufficiently long time for moving out from the hot region to reach high *sp*-carbon concentration but short enough to avoid further decomposition of wires into atoms and dimers. This characteristic time is dependent on the temperature, thus indicating a complex mechanism that is difficult to optimize by tuning depositon conditions.

Once formed, *sp*-carbon wires must be deposited on the substrate preventing fragmentation and reorganization. A substrate position far with respect to the shock front is also relevant to reduce the kinetic energy of species in the gas phase. Low kinteic energy depositon regime is similar to the one occurring in low energy supersonic cluster beam deposition, as shown in experimental works and numerical simulations [47,48]. Under these optimal conditions the as produced material is a *sp-sp²* moiety in which CAWs appear to be embedded in an amorphous mainly $sp^2$ carbon network [49].

Due to the high porosity the film density is expected to be very low. In fact, based on other works using PLD for the depositon of carbon foams we can estimate that the density of the material produced at 500 Pa Ar is well below 0.01 g/cm$^3$ [50,51].

The stability of CAWs in air is up to a few days and longer when samples are stored in vacuum (data not shown). In this case the use of SERS-active substrates is beneficial since silver nanoparticles can help in improving the stability of CAWs, as we have outlined in the case of H-terminated polyynes embedded in silver nanoparticle assemblies [52].

Regarding the structure of CAWs produced with our approach, it is reasonable to assume that CAWs are terminated by *sp*$^2$ carbon in many different ways. This can include both double bond and triple bond terminations which are able to induce both polyyne-like and cumulene-like wires [53]. At the same time distorsions from the linear structure (i.e. bending and torsions) are expected according to the highly disordered network. All these effects can modify the vibrational features making it difficult to unveil the detailed structure of wires. In addition SERS effect can introduce further modifications to the spectra due to charge transfer effects activated by the interaction with silver clusters of the SERS active medium. In fact it has been shown that the features in SERS spectra of polyynes in solution are red-shifted with respect to those in Raman spectra, as a result of the interaction with noble metal nanoparticles [31, 34]. This occurrence is compatible with the relatively low frequency observed in our SERS spectra of the main sp carbon peak (i.e. at about 1900 cm$^{-1}$) which for polyyne is usually above 2000 cm$^{-1}$. Even though structural distorsions and size-effects can change the vibrational features, peaks below 2000 cm$^{-1}$ are typical of cumulene-like structures. To this respect charge transfer from metal to wires would correspond to a decrease of the bond length alternation (BLA) thus indicating a transition towards more equalized wires, i.e. with cumulene-like structure [34]. In conclusion, the large width of the overall *sp*-carbon band is due to a convolution of modes coming from a wide distribution in size, termination, torsional and bending effects that is compatible with the presence of many different CAWs with both polyyne-like and cumulene-like organization and with different size [54].

## 4. Conclusions

We have shown that nanosecond PLD in the presence of a background gas allows the synthesis of *sp*-carbon-atom wires embedded or mixed in mainly *sp*$^2$ amorphous carbon films. The use of SERS active substrates permits to detect the presence of *sp*-carbon phase and to investigate the role of deposition parameters in the formation of carbon-atom wires. Results suggest that carbon-atom wires form in the gas phase due to the strong out-of-equilibrium conditions occurring in the plasma plume. Background gas plays a crucial role both in the plume confinement and in slowing down the species before reaching the substrate surface, as evidenced by the effect of changing the substrate distance from the target. Our results open the way to the controlled deposition of hybrid *sp-sp$^2$* films through the exploitation of fine control of ns-PLD on the film growth mechanisms. By tuning of the deposition process the morphology, density and structure (including hybridization state) of carbon films can be controlled in a wide range. Thin films of *sp-sp$^2$* moieties with open porous morphology and with high effective surface are optimal candidates for potential applications. As an example theoretical predictions have shown that a sp carbon network decorated by Ca atoms is an optimal candidate material for hydrogen storage applications [55]. In this context, the versatility of PLD technique in the deposition of different materials (e.g. metals, oxides, polymers), including the possibility to control doping, opens new perspectives for the synthesis of carbon-atom wires with tailored functionality in view of future applications.